\documentstyle[twocolumn,seceq]{jpsj}
\input epsf.tex

\def\runtitle{Spin-Orbital Wave Excitations in Orbitally Degenerate Exchange Model with Multipolar Interactions}
\def\runauthor{Hiroaki {\sc Kusunose} and Yoshio {\sc Kuramoto}}

\title
{Spin-Orbital Wave Excitations in Orbitally Degenerate Exchange Model with Multipolar Interactions}

\author
{
Hiroaki {\sc Kusunose}\footnote{E-mail: kusu@cmpt.phys.tohoku.ac.jp}
and
Yoshio {\sc Kuramoto}
}

\inst
{
Department of Physics, Tohoku University, Sendai 980-8578
}

\recdate
{
April 16, 2001
}

\abst
{
Elementary excitations in the multipole ordered state, which models the phase III in CeB$_6$, are investigated by means of a generalized Holstein-Primakoff formalism.
When different kinds of nearest-neighbor exchange interactions between multipoles are comparable to each other, orbital-flip excitations exhibit almost one-dimensional dispersion along the $z$ axis.
With high symmetry of the interactions, zero modes appear due to the macroscopic degeneracy of the ground state.
The next-nearest-neighbor dipole-dipole interaction, which stabilizes the magnetic order of the phase III in CeB$_6$, lifts the degeneracy and leads to gapfull excitation spectrum.
When the octupole-octupole next-nearest-neighbor interaction exists simultaneously, the spectrum shows softening at $\Gamma$ and Z points.
These excitations may be probed by neutron scattering and ultrasonic measurements.
}

\kword
{
orbital degeneracy, multipole moment, Holstein-Primakoff transformation, CeB$_6$
}

\begin{document}
\sloppy
\maketitle

%
%
\section{Introduction}
Among many fascinating compounds with orbital degeneracy, CeB$_6$ shows an interesting entanglement of dipole, quadrupole as well as octupole moments\cite{takigawa83,effantin85,erkelens87,fujita80,shiina97,uimin96,kuramoto98}.
In particular, the entanglement gives a natural explanation for the longstanding inconsistency between the neutron diffraction and the B$^{11}$-NMR measurements of the phase II\cite{sakai98}.
Namely, the field-induced octupole moment of $\Gamma_2$-type, which arises through the mode mixing with the dipole moment, causes an extra internal field on the B nucleus sites.
It explains the unexpected field-angle dependence in the NMR measurement.

In discussing the transition from the antiferro-quadrupolar (AFQ) phase, called II, to the magnetically ordered phase III, the mode mixing between dipoles and octupoles also plays an important role\cite{sera99,shiba99}.
The present authors have made systematic analysis of multipole modes under the AFQ ordering\cite{kusunose01}.
Then they have discussed effects of multipolar interactions on magnetic phases within the mean-field approximation.
In their model Hamiltonian, the pseudo-dipole-type interactions for the next-nearest neighbors (NNN) are indispensable to account for the observed non-collinear magnetic ordering.
The usual exchange-type interactions for the nearest neighbors (NN) are also important.

What they found are the followings\cite{kusunose01}: (i) the competition between the NN dipole-dipole and octupole-octupole interactions strongly suppresses transition temperatures for simple AF magnetic order, (ii) the ferromagnetic NNN dipole-dipole interaction plays a dominant role in stabilizing the non-collinear ordering in the phase III, and (iii) the AF NNN octupole-octupole interaction is expected to be as strong as the dipole-dipole interaction since the phase III$^\prime$ is observed for relatively weak magnetic field\cite{effantin85,erkelens87}.

For magnitude of the coupling constants in the model Hamiltonian, a reliable first-principle calculation is not available at present.
However, the nature of the interaction parameters should be reflected in characteristics of elementary excitations.
The purpose of the present paper is to derive spectra of all elementary excitations and to investigate how they reflect the details of interactions.
The analysis also provides us with a typical example of collective excitations for systems where both the spin and the orbital degrees of freedom are involved\cite{thalmeier98,joshi99}.
In this respect, a comparison with orbital ordered manganites\cite{ishihara97,khaliullin97,brink98,maezono00} would be useful, in which weak coupling between spin and orbital degrees of freedom gives rise to independent spin and orbital waves.
In the present case, the excitations all involve magnetic moments due to the strong spin-orbit coupling of $f$ electrons.
Therefore all excitations can be measured by neutron scattering measurement.
Some low-lying modes may also couple with ultrasonic waves.
An experimental study for CeB$_6$\cite{bouvet93} indicates strange structure of low-lying excitations.
Detailed experimental studies would be helpful to determine interaction parameters in the model Hamiltonian.

This paper is organized as follows.
In \S2 we set up the model Hamiltonian for CeB$_6$\cite{shiina97,shiba99,kusunose01}.
Then, a generalized Holstein-Primakoff formalism is developed by taking the reference state as the mean-field ground state\cite{onufrieva85,joshi99}.
In \S3 the dispersion relations of spin-orbital wave excitations are given for some typical parameters in question.
The summary is given in the final section.
Details of the Holstein-Primakoff transformation are explained in the Appendix.

%
%
\section{Model and Formalism}
\subsection{Model Hamiltonian}
The four states in the $\Gamma_8$ level, which is the crystalline-electric-field ground state of Ce$^{3+}$ ion, are represented in terms of the basis $|J_z\rangle$ of $J=5/2$ as follows:
\begin{equation}
\left|1\pm\right)=\sqrt{\frac{5}{6}}\left|\pm\frac{5}{2}\right\rangle+\sqrt{\frac{1}{6}}\left|\mp\frac{3}{2}\right\rangle,\;\;\left|2\pm\right)=\left|\pm\frac{1}{2}\right\rangle.
\end{equation}
The entanglement of spin and orbital degrees of freedoms can be described most physically by multipole operators in accordance with the point-group symmetry\cite{shiina97}.
The multipole moments are expressed in a concise way by means of two Pauli matrices, ${\mib\sigma}$ and ${\mib\tau}$.
Here ${\mib\tau}$ acts on the orbital partners, while ${\mib\sigma}$ on the Kramers pairs, and is called spin hereafter.
To specify a multipole operator, an irreducible representation $\Gamma$ with possible multiplicity, and a component $\gamma$ are required.
The set ($\Gamma$, $\gamma$) is abbreviated as $A$.
In the irreducible representation $\Gamma$, the subscript $u$ represents the odd property under the time reversal, and $g$ the even one.
The explicit form of the operators $X^A$ are summarized in Table I, where we have introduced linear combinations of $\tau^x$ and $\tau^z$ as
\begin{equation}
\eta^\pm=\frac{1}{2}\left(\pm\sqrt{3}\tau^x-\tau^z\right),\;\;
\zeta^\pm=-\frac{1}{2}\left(\tau^x\pm\sqrt{3}\tau^z\right).
\end{equation}

Experimentally, the $\Gamma_{5g}$-type quadrupolar ordering (phase II) first occurs at ${\mib Q}=(1/2,1/2,1/2)$ in units of $2\pi/a$, where $a$ is the lattice constant.
As temperature decreases, the non-collinear magnetic ordering (phase III) is realized.
The magnetic ordering is coexistent with the AFQ ordering, which is characterized by the double ${\mib k}$ modulation vectors, i.e., ${\mib k}_1=(1/4,1/4,1/2)$ and ${\mib k}_2=(1/4,-1/4,1/2)$\cite{effantin85,erkelens87}.

This complicated magnetic structure can be understood by the fact that a lowering of the point-group symmetry brought about by the AFQ ordering mixes otherwise independent multipole modes\cite{kusunose01}.
For example, in the case of $O_{xy}$ order the following mode mixings arise:
\begin{eqnarray}
&& {\rm (i)}\;\;\; \Gamma_{2u} {\rm -} \Gamma_{4u1}(z), \\
&& {\rm (ii)}\;\;\; \Gamma_{4u2}(x,y) {\rm -} \Gamma_{5u}(x,y),
\end{eqnarray}
where $\Gamma_{4u1}$ and $\Gamma_{4u2}$ represent $\Gamma_{4u}$ with multiplicity index 1 or 2.
Since the order parameters in (ii) are concerned with realization of the phase III, it is natural to expect that the $\Gamma_{4u2}$-type dipole-dipole and the $\Gamma_{5u}$-type octupole-octupole interactions are the most important.

We assume that the intersite interactions between $i$ and $j$ are classified according to the representations of the cubic group.
The conduction electrons that give rise to the Kondo effect are not treated explicitly\cite{kusunose99a,kuramoto98a,kusunose99b}.
Then the Fourier transform of the NN interaction for the 3D simple cubic lattice is given by
\begin{equation}
J_{\Gamma}({\mib q})_{\gamma\gamma'}=-J_{\Gamma}\delta_{\gamma\gamma'}(\cos{q_x}+\cos{q_y}+\cos{q_z})/3.
\label{nnint}
\end{equation}
To realize the AFQ ordering of $\Gamma_{5g}$-type, $J_{5g}$ should be positive and the largest among all interactions, and hence we denote $J_{5g}=T_Q$.
The strength of the NN interactions mediated by conduction electrons was discussed\cite{shiba99,sera99,ohkawa83} on the basis of the group-theoretical argument, where it was concluded that there are only two independent couplings.

In order to account for the observed non-collinear spin orientation, we introduce the NNN interaction $K$ of the pseudo-dipole type\cite{kusunose01,sakai}:
\begin{equation}
K^{\Gamma\gamma\gamma'}_{ij}=-K_{\Gamma}(\delta^{\gamma\gamma'}-3n^\gamma_{ij} n^{\gamma'}_{ij})/12,
\end{equation}
where ${\mib n}_{ij}$ is the unit vector across $i$ and $j$ sites.
The pseudo-dipole interaction is considered for three-dimensional odd representations, i.e., $\Gamma_{4u1}$, $\Gamma_{4u2}$ and $\Gamma_{5u}$.
Note that the positive (negative) $K_{\Gamma}$ favors (anti) parallel alignment along ${\mib n}_{ij}$.
In the {\mib q} space this interaction is written as
\begin{equation}
K_{\Gamma}({\mib q})_{\gamma\gamma'}=-K_{\Gamma}
\left(\begin{array}{ccc}
j_x & j_{xy} & j_{zx} \\
j_{xy} & j_y & j_{yz} \\
j_{zx} & j_{yz} & j_z
\end{array}\right)_{\gamma\gamma'},
\label{pdint}
\end{equation}
where
\begin{eqnarray}
&&j_x=\frac{1}{6}\biggl[2\cos(q_y)\cos(q_z)-\cos(q_z)\cos(q_x)\nonumber\\
&&\;-\cos(q_x)\cos(q_y)\biggr],\;
j_{xy}=\frac{1}{2}\sin(q_x)\sin(q_y),
\end{eqnarray}
and other components are given by cyclic rotation of $q_x$, $q_y$, and $q_z$.
The microscopic origin of the pseudo-dipole type interactions has been discussed by taking into account the RKKY or the superexchange interactions together with the $d$-$f$ exchange or the Hund's-rule coupling\cite{sakai,kasuya66,yildrim95}.

Consequently, the Hamiltonian used in this paper is given by
\begin{eqnarray}
H&&=-\frac{1}{2}\sum_{i\ne j}\sum_{AB}D^{AB}_{ij}X^A_iX^B_j \nonumber \\
&&=-\frac{1}{2}\sum_{\mibs q}\sum_{AB}D^{AB}_{\mibs q}X^A_{\mibs q}X^B_{-{\mibs q}},
\label{exham}
\end{eqnarray}
with the interaction
\begin{equation}
{\sf D}_{\Gamma}({\mib q})=\left\{
\begin{array}{ll}
{\sf J}_{\Gamma}({\mib q}), & (\Gamma=2u,\;3g,\;5g),\\
{\sf J}_{\Gamma}({\mib q})+{\sf K}_{\Gamma}({\mib q}), & (\Gamma=4u1,\;4u2,\;5u).
\end{array}\right.
\end{equation}

\begin{figure}
\begin{center}
\epsfxsize=8cm \epsfbox{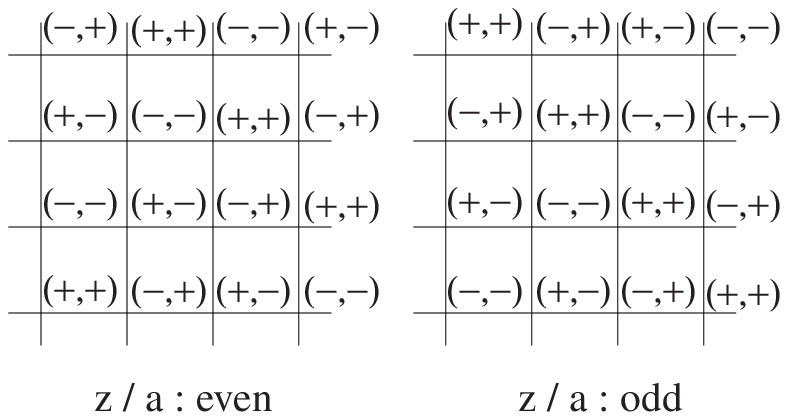}
\end{center}
\caption{The real-space arrangement of the phase III in the (001) plane with the coordinate $z$. The symbol $(\xi_i,\mu_i)$ characterize the quadrupole sublattice ($\xi_i$) and the spin direction ($\mu_i$) along the easy axis of each quadrupole sublattice.}
\label{ximupat}
\end{figure}

According to the mean-field analysis\cite{sera99,shiba99,kusunose01}, the transition temperature of the simple AF ordering is suppressed by the competitions between $J_{2u}$ and $J_{4u1}$ and between $J_{4u2}$ and $J_{5u}$.
In the case of the perfect competition, i.e., $J_{2u}=J_{4u1}$ and $J_{4u2}=J_{5u}$, the observed non-collinear ordering is stabilized for $|K_{5u}|/K_{4u2}<1$ with $K_{4u2}>0$\cite{kusunose01}.
The corresponding order parameters at $T=0$ are given by
\begin{eqnarray}
&& \langle {\mib X}^{4u2}_i \rangle_{\rm MF} = A_+\mu_i(\xi_i,-1,0)^t/\sqrt{2}, \nonumber \\
&& \langle {\mib X}^{5u}_i \rangle_{\rm MF} = A_-\mu_i(\xi_i,1,0)^t/\sqrt{2}, \nonumber \\
&& \langle X^{5gz}_i \rangle_{\rm MF} = \xi_i,
\label{ops}
\end{eqnarray}
where we have defined $A_\pm=(\sqrt{3}\pm1)/2$ and $(a,b,c)^t$ denotes a column vector.
The site dependences are expressed by
\begin{eqnarray}
&& \xi_i=\exp(i{\mib Q}\cdot {\mib R}_i), \\
&& \mu_i=\frac{1}{\sqrt{2}}\bigl[
\cos({\mib k}_1\cdot{\mib R}_i+\frac{\pi}{4})
+\cos({\mib k}^\prime_1\cdot{\mib R}_i-\frac{\pi}{4})
\nonumber \\ && \mbox{\hspace{5mm}}
-\cos({\mib k}_2\cdot{\mib R}_i+\frac{\pi}{4})
+\cos({\mib k}^\prime_2\cdot{\mib R}_i-\frac{\pi}{4})
\bigr],
\end{eqnarray}
which specify the quadrupole sublattice ($\xi_i=\pm1$) and the spin direction ($\mu_i=\pm1$) along the easy axis of each quadrupole sublattice.
Here we have defined ${\mib k}_n'={\mib Q}-{\mib k}_n$ ($n=1,2$).
The real-space arrangement of $(\xi_i,\mu_i)$ is shown in Fig. \ref{ximupat}.
These mean-field solutions can be used as the reference state of elementary excitations in the Holstein-Primakoff formalism as will be developed in the next subsection.

\subsection{Generalized Holstein-Primakoff transformation}
For the purpose of investigating the elementary excitations, we first develop a generalized Holstein-Primakoff transformation for the Hamiltonian, (\ref{exham}).

Let us denote the basis state for each site $i$ by $|n\rangle$, where the quantization axes are common with all sites.
Note that the multipole operators listed in Table I have been defined in this coordinate system.
In the presence of a long-range order, especially a non-collinear one, it is useful to introduce a local coordinate $|n\rangle_i$, which is taken according to be the ordering direction of the classical multipole.
The local coordinate has been indicated by the subscript $i$ of the ket vector.
Practically, this can be done by the site-dependent unitary transformation,
\begin{equation}
|n\rangle=\sum_{m} U^i_{mn} |m\rangle_i.
\label{lu}
\end{equation}
In the case of the AF Heisenberg model for instance, such unitary transformation is given by the rotation $\pi$ with respect to the $x$ axis on one of two sublattices.
The unitary transformation is defined as such that the resultant local state $|n\rangle_i$ becomes an eigenstate of the local mean-field Hamiltonian,
\begin{equation}
H^{\rm MF}_i=-\sum_A {\hat X}^A_i\phi^A_i,
\label{mfham}
\end{equation}
and its eigenvalues $\epsilon_n$ are common to all sites.
The averages of the physical operators with respect to the ground state reproduce the mean-field solutions, i.e., $\langle 0 |{\hat X}^A_i| 0 \rangle_i=\langle {\hat X}^A_i \rangle_{\rm MF}$.

We introduce the Holstein-Primakoff bosons, $b^\dagger_{ni}$, which describe local excitations from the ground state $|0\rangle_i$ to excited states $|n\rangle_i$.
Since we work with the local coordinate, the structure of the local excitations is common for all sites.
In terms of these bosons, the Hamiltonian (\ref{exham}) is expressed (see Appendix for details) up to the second order as
\begin{eqnarray}
&&
{\cal H}\equiv H-\frac{1}{2}NM^{0000}_{00}({\mib 0})+\sum_{\mibs k}{}^\prime\sum_m\sum_{M}\Omega^{mm}_{MM}({\mib k}) \nonumber\\
&&\mbox{\hspace{2mm}}
=\frac{1}{2}\sum_{\mibs k}{}^\prime\sum_{mn}\sum_{MN}
(b_{m,-M}(-{\mib k})\;\;b^\dagger_{m,M}({\mib k}))
\times\nonumber\\&&\mbox{\hspace{1cm}}\times
\left(\begin{array}{cc}
\Omega^{mn}_{MN}({\mib k}) & \Lambda^{mn}_{MN}({\mib k}) \\
\Lambda^{mn}_{MN}({\mib k}) & \Omega^{mn}_{MN}({\mib k})
\end{array}\right)
\left(\begin{array}{c}
b^\dagger_{n,-N}(-{\mib k}) \\ b_{n,N}({\mib k})
\end{array}\right),\nonumber\\
\label{boham}
\end{eqnarray}
where the summation over ${\mib k}$ is restricted to the magnetic Brillouin zone, i.e., ${\mib k}$ is regarded as the reduced wave vector.
Here we have defined the bosons in the ${\mib k}$ space with the band suffix $M$ as
\begin{equation}
b^\dagger_{m,M}({\mib k})=b^\dagger_{m,{\mibs k}+{\mibs G}_M},
\end{equation}
where the reciprocal lattice vectors ${\mib G}_M$ are defined as such that ${\mib k}+{\mib G}_M$ spans the original Brillouin zone, and ${\mib G}_0={\mib 0}$.
The matrix elements of $\Omega$ and $\Lambda$ are given by
\begin{eqnarray}
&&\Omega^{mn}_{MN}({\mib k})=M^{00mn}_{0,N-M}({\mib 0})+M^{m00n}_{M,N}({\mib k})-\delta_{mn}M^{0000}_{0,N-M}({\mib 0}), \nonumber \\
&&\Lambda^{mn}_{MN}({\mib k})=M^{0m0n}_{M,N}({\mib k}),
\label{summatelm}
\end{eqnarray}
with
\begin{equation}
M^{mnkl}_{M,N}({\mib k})=\sum_{AB}\sum_{K}D^{AB}_{{\mibs k}+{\mibs G}_K}(X^A_{{\mibs G}_M-{\mibs G}_K})_{mn}(X^B_{{\mibs G}_K-{\mibs G}_N})_{kl}.
\label{matelm}
\end{equation}
Note that the ground state energy except for the zero-point energy has been subtracted from the Hamiltonian.

By the following Bogoliubov transformation:
\begin{equation}
\alpha_s({\mib k})=\sum_m\sum_M\bigl[u^s_{mM}({\mib k})b_{m,-M}(-{\mib k})+v^s_{mM}({\mib k})b^\dagger_{mM}({\mib k})\bigr],
\end{equation}
the Hamiltonian can be diagonalized as
\begin{equation}
{\cal H}=\sum_{\mibs k}{}^\prime\sum_s \omega_s({\mib k})\biggl(\alpha^\dagger_s({\mib k})\alpha_s({\mib k})+\frac{1}{2}\biggr).
\end{equation}
The dispersion relation $\omega_s({\mib k})$ and the coefficients $u^s$ and $v^s$ are obtained by the eigenvalue equation,
\begin{eqnarray}
&&
|\omega-F^{mn}_{MN}({\mib k})|=0, \label{evp} \\
&&
F^{mn}_{MN}({\mib k})=
\left(\begin{array}{cc}
\Omega^{mn}_{MN}({\mib k}) & -\Lambda^{mn}_{MN}({\mib k}) \\
\Lambda^{mn}_{MN}({\mib k}) & -\Omega^{mn}_{MN}({\mib k})
\end{array}\right).
\end{eqnarray}

\subsection{Local coordinate in the case of CeB$_6$}
In order to rotate to the ordering direction, let us introduce the following unitary transformation,
\begin{equation}
U^i=\exp(-i\pi\mu_i\sigma^x_i/4)\exp(-i\pi\xi_i\sigma^z_i/8)Q\exp(-i\pi\xi_i\tau^x_i/4),
\end{equation}
with
\begin{equation}
Q=\left(\begin{array}{cccc}
1 & 0 & 0 & 0 \\
0 & 0 & 0 & 1 \\
0 & 0 & 1 & 0 \\
0 & 1 & 0 & 0
\end{array}\right).
\end{equation}
As will be shown shortly, exponential operators rotate spin and orbital states so that $H_i^{\rm MF}$ contains diagonal components $\sigma^z_i$ and $\tau^z_i$ only.
The matrix $Q$ interchanges the second and forth spin-orbital states.
Then, the multipole operators are transformed to
\begin{eqnarray}
&&X^{2u}_i=-\mu_i\xi_i\tau^z_i\sigma^y_i, \nonumber \\
&&{\mib X}^{3g}_i=\biggl[\mu_i\xi_i\tau^y_i\sigma^y_i, \tau^x_i\biggr], \nonumber \\
&&{\mib X}^{4u1}_i=\biggl[\sigma^+_i\tau^x_i, -\xi_i\sigma^-_i\tau^x_i, -\mu_i\sigma^y_i\biggr], \nonumber \\
&&{\mib X}^{4u2}_i=\biggl[\frac{1}{2}(\sqrt{3}\sigma^+_i-\sigma^-_i\tau^z_i), \frac{\xi_i}{2}(\sqrt{3}\sigma^-_i-\sigma^+_i\tau^z_i), -\xi_i\tau^y_i\biggr], \nonumber \\
&&{\mib X}^{5u}_i=\biggl[-\frac{1}{2}(\sigma^+_i+\sqrt{3}\sigma^-_i\tau^z_i), \frac{\xi_i}{2}(\sigma^-_i+\sqrt{3}\sigma^+_i\tau^z_i), -\mu_i\tau^x_i\sigma^y_i\biggr], \nonumber \\
&&{\mib X}^{5g}_i=\biggl[\sigma^-_i\tau^y_i, \xi_i\sigma^+_i\tau^y_i, \xi_i\tau^z_i\biggr],
\label{mulop}
\end{eqnarray}
where we have defined $\sigma^\pm_i=(\sigma^x_i\pm\mu_i\xi_i\sigma^z_i)/\sqrt{2}$.

\begin{figure}
\begin{center}
\epsfxsize=6cm \epsfbox{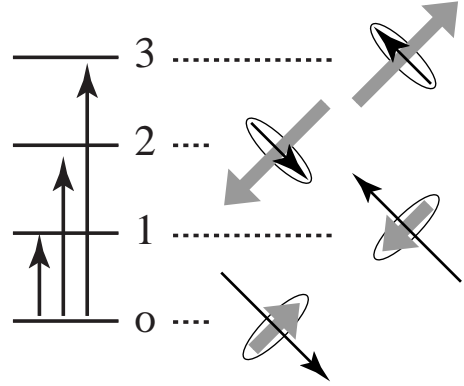}
\end{center}
\caption{The schematic picture of local excitations. The ovals represent the $O_{xy}$ quadrupole, thin arrows the $\Gamma_{4u2}$-type dipole, and thick arrows the $\Gamma_{5u}$-type octupole moments.}
\label{localex}
\end{figure}

We factorize the mean fields in eq.~(\ref{mfham}) into its order parameter, eq.~(\ref{ops}), and coupling strength, $\lambda_\Gamma$ as
\begin{eqnarray}
&&{\mib\phi}^{4u2}_i=\langle{\mib X}^{4u2}_i\rangle_{\rm MF}\lambda_{4u2}, \\
&&{\mib\phi}^{5u}_i=\langle{\mib X}^{5u}_i\rangle_{\rm MF}\lambda_{5u}, \\
&&\phi^{5gz}_i=\langle X^{5gz}_i\rangle_{\rm MF}\lambda_{5g}.
\end{eqnarray}
With $\xi_i^2=\mu_i^2=1$, the mean-field Hamiltonian is diagonalized as
\begin{equation}
H^{\rm MF}_i=-A_+\frac{\sqrt{3}+\tau^z_i}{2}\sigma^z_i\lambda_{4u2}
-A_-\frac{\sqrt{3}\tau^z_i-1}{2}\sigma^z_i\lambda_{5u}-\tau^z_i\lambda_{5g},
\end{equation}
which depends neither on the sign of $\xi_i$ nor that of $\mu_i$ as desired.
The eigenstates are specified by $|\tau_z,\sigma_z\rangle_i$, i.e., $|0\rangle_i\equiv|+,+\rangle_i$, $|1\rangle_i\equiv|+,-\rangle_i$, $|2\rangle_i\equiv|-,+\rangle_i$ and $|3\rangle_i\equiv|-,-\rangle_i$, and their energies are given by
\begin{eqnarray}
&&\epsilon_{+\pm}=\mp(A_+^2\lambda_{4u2}+A_-^2\lambda_{5u})-\lambda_{5g}, \\
&&\epsilon_{-\pm}=\mp A_+A_-(\lambda_{4u2}-\lambda_{5u})+\lambda_{5g}.
\end{eqnarray}
The schematic picture of local excitations is shown in Fig. \ref{localex}, in which the oval represents the $O_{xy}$ quadrupole and the thin (thick) arrow does the $\Gamma_{4u2}$-type dipole (the $\Gamma_{5u}$-type octupole).
Note that the dipole and the octupole moments change their magnitude in the excitation processes with orbital flip $(n=2,3)$, while they do not in the spin-flip process $(n=1)$.

By using the matrix elements of multipole operators, (\ref{mulop}) for eqs.~(\ref{matelm}) and (\ref{summatelm}), the dispersion relation is calculated by eq.~(\ref{evp}).

\section{Spin-Orbital Wave Excitations}
\begin{figure}
\begin{center}
\epsfxsize=6cm \epsfbox{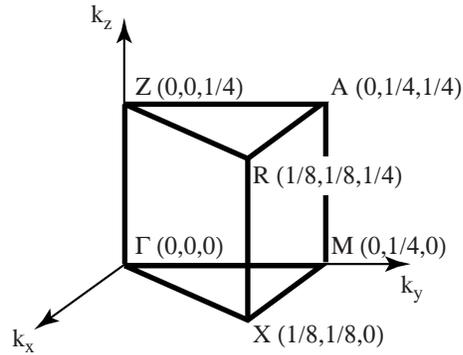}
\end{center}
\caption{The high-symmetry points in the magnetic Brillouin zone with tetragonal symmetry.}
\label{magbz}
\end{figure}

In the phase III, the magnetic unit cell has the tetragonal symmetry and is 16 times larger than the paramagnetic one.
Thus, there are $16\times 3=48$ modes in the magnetic Brillouin zone.
The definition of the high-symmetry points of the magnetic Brillouin zone is given in Fig. \ref{magbz}.

Let us first investigate the case of the NN interactions only, i.e., $K_\Gamma=0$.
Figure 4(a) shows the case of the SU(4) limit, i.e., $J_\Gamma=T_Q$ for all $\Gamma$.
The dispersion relations are completely one dimensional; $\omega$ varies only along the $z$ axis\cite{joshi99}.
There are 16 zero modes, which involve only the spin-flip ($n=1$) excitations.
Since without the orbital flip, the dipole-dipole and the octupole-octupole interactions cancel with each other\cite{kusunose01}, the ground state has a macroscopic degeneracy, $2^N$, which constitutes the zero modes.
Namely, the assumed non-collinear ordering is never stabilized in the case of the perfect competition.
On the other hand, the excitations with $n=2, 3$ bosons are always accompanied with the orbital flip, and the corresponding collective modes have the finite energy.

When we go away from the SU(4) limit as shown in Fig.~4(b), some modes acquire dispersion in the $\Gamma$-X-M plane.
However, overall structure of excitations does not change.
There still remains almost one-dimensional excitations.
When we set $J_{3g}=J_{4u2}=J_{5u}=0$ and leave only the competition between $J_{2u}$ and $J_{4u1}$, the bunch of excitations around $\omega/T_Q\sim 1$ and $2$ in Fig.~4(b) merge roughly into one with $\omega/T_Q\sim 2$ as shown in Fig.~4(c).
If we leave the other competition, $J_{4u2}=J_{5u}=0.9T_Q$, two bunches of modes in Fig.~4(b) split into four bunches along Z-A-R-Z axis ($\omega/T_Q\sim 1, 1.5, 2$ and $2.5$ as shown in Fig.~4(d)).

In any case, orbital-flip excitations with $\omega$ being larger than $T_Q$ show almost one-dimensional dispersion relations.
The spin-flip excitations constitute all zero modes in consequence of the perfect competition between the NN interactions and of the lack of the NNN interactions.
If we discuss possible lifting of the macroscopic degeneracy without the NNN interactions, we should include quantum fluctuations in the ground state.
Such consideration is out of the scope of this study.

Next, we consider the effect of the NNN interactions with the NN interactions being fixed close to the SU(4) limit, i.e., $J_{2u}=J_{3g}=J_{4u1}=J_{4u2}=J_{5u}=0.9T_Q$.
As $K_{4u2}$ is switched on and increased, the zero modes go up without losing dispersionless character and they form Ising-like excitations with a gap of the order of $2K_{4u2}$.
The case of $K_{4u2}=0.5T_Q$ is shown in Fig.~5(a).
The Ising-like excitations can be understood in terms of the strong anisotropy with respect to the rotations of the dipole and the octupole moments in the presence of the quadrupolar ordering.

In order to explain the stability of the phase III$^\prime$ in magnetic field, it has been proposed that the AF octupole-octupole interaction, $K_{5u}$, should be as strong as the dipole-dipole interaction, i.e., $K_{5u}\sim-K_{4u2}$\cite{kusunose01}.
Such situation in the coupling constants suggests that the $\Gamma_{5u}$-type octupole is a reasonable candidate for the order parameter in the phase IV, which was found recently in Ce$_x$La$_{1-x}$B$_6$ with $x\sim 0.75$\cite{nakamura97,hiroi97,tayama97,kohgi,kuramoto00}.
In the case of $K_{4u2}=0.5T_Q$ and $K_{5u}=-0.45T_Q$, the Ising-like excitations become quite dispersive and show softening at $\Gamma$ and $Z$ points as shown in Fig.~5(b).
If one increases $|K_{5u}|$ beyond $K_{4u2}$, the energy of the soft modes becomes complex.
This means that the assumed ground state is unstable against exponentially growing modes.

\section{Summary}
We have investigated the spin-orbital wave excitations for the phase III of CeB$_6$.
Using the mean-filed solution as the classical ground state, we calculate the dispersion relations in a generalized Holstein-Primakoff formalism.

When the nearest-neighbor dipole-dipole and octupole-octupole interactions are comparable, the dispersion of orbital-flip excitations is almost one-dimensional along the $z$ axis.
The spin-flip excitations constitute zero modes owing to the macroscopic degeneracy of the ground state.

As the next-nearest-neighbor dipole-dipole interaction which stabilizes the magnetic order of the phase III increases, the zero modes go up keeping the dispersionless character and form gapfull Ising-like excitation spectrum.
When the dipole-dipole and the octupole-octupole interactions exist simultaneously, the Ising-like excitation shows softening at $\Gamma$ and Z points.
We expect that these modes can be probed experimentally.
Detailed measurements of excitations would reveal the nature of spin-orbital excitations and relevant intersite interactions.

%
%
\section*{Acknowledgements}
H. K. would like to thank H. Shiba for stimulating discussions on the orbital waves.
He has also benefited from fruitful conversations with N. Fukushima and T. Matsumura.
The authors thank T. Goto for his comment on possible coupling with ultrasound.

%
%
\appendix
\section{Details of Generalized Holstein-Primakoff Transformation}
In this appendix, detailed derivation of (\ref{boham}) is given by introducing a generalized Holstein-Primakoff transformation for the Hamiltonian, (\ref{exham}).

From now on we work with the local coordinate which is defined by (\ref{lu}).
By introducing the Hubbard operator at $i$ site, $S^{mn}_i=|m\rangle\langle n|_i$, the Hamiltonian, (\ref{exham}), is expressed as
\begin{equation}
H=\frac{1}{2}\sum_{ij}\sum_{mnkl}M^{mnkl}_{ij}S^{mn}_iS^{kl}_j,
\end{equation}
where the intersite interactions are given in terms of the matrix elements of the multipole operators.
In the local coordinate we obtain
\begin{equation}
M^{mnkl}_{ij}=D^{AB}_{ij}\left(X^A_i\right)_{mn}\left(X^B_j\right)_{kl}.
\end{equation}

The generalized Holstein-Primakoff transformation is defined as (for $m$, $n\ne 0$)
\begin{eqnarray}
&&S^{00}_i=P-\sum_\ell{}^\prime b^\dagger_{\ell i}b_{\ell i}, \\
&&S^{m0}_i=b^\dagger_{m i}\;\sqrt{P-\sum_\ell{}^\prime b^\dagger_{\ell i}b_{\ell i}}, \\
&&S^{0m}_i=\sqrt{P-\sum_\ell{}^\prime b^\dagger_{\ell i}b_{\ell i}}\;\; b_{mi}, \\
&&S^{mn}_i=b^\dagger_{mi}b_{ni},
\end{eqnarray}
where the prime means that the summation excludes $\ell=0$.
The commutation relation for the Hubbard operators,
\begin{equation}
[S^{mn}_i,S^{kl}_j]=(\delta_{nk}S^{ml}_i-\delta_{lm}S^{kn}_i)\delta_{ij},
\end{equation}
is satisfied if one uses boson commutation relations for $b_{\ell i}$ and $b^\dagger_{\ell i}$.
The boson creation operator $b^\dagger_{\ell i}$ represents the local excitation from the ground state to the state $\ell$.
The symbol $P$ is the expectation value of $S^{00}_i$ with respect to the vacuum of bosons.
Then we can perform a systematic expansion of the Hamiltonian in power of $1/P$.
After the expansion we set $P=1$ for the physical system.

Up to ${\cal O}(1/P)$, the Hamiltonian is expressed as
\begin{eqnarray}
&&H=\frac{1}{2}\biggl[NM^{0000}_{{\mibs 0},{\mibs 0}}+\sqrt{N}\sum_{\mibs k}\sum_m{}^\prime\bigl(M^{m000}_{{\mibs k},{\mibs 0}}b^\dagger_{m,{\mibs k}}+{\rm h.c.}\bigr)\biggr] \nonumber \\
&&
+\sum_{{\mibs k}{\mibs p}}\sum_{mn}{}^\prime\biggl[\Omega^{mn}_{{\mibs k},{\mibs p}}b^\dagger_{m,{\mibs k}}b_{n,{\mibs p}}+\frac{1}{2}\bigl(\Lambda^{mn}_{{\mibs k},{\mibs p}}b_{m,-{\mibs k}}b_{n,{\mibs p}}+{\rm h.c.}\bigr)\biggr], \nonumber\\&&
\label{bh0}
\end{eqnarray}
where $N$ is the number of sites and the Fourier transform of $M^{mnkl}_{ij}$ is given by
\begin{equation}
M^{mnkl}_{{\mibs k},{\mibs p}}=\sum_{\mibs q}D^{AB}_{\mibs q}(X^A_{{\mibs k}-{\mibs q}})_{mn}(X^B_{{\mibs q}-{\mibs p}})_{kl}.
\end{equation}
Here we have defined
\begin{eqnarray}
&&\Omega^{mn}_{{\mibs k},{\mibs p}}=-\delta_{mn}M^{0000}_{{\mibs 0},{\mibs p}-{\mibs k}}+M^{00mn}_{{\mibs 0},{\mibs p}-{\mibs k}}+M^{m00n}_{{\mibs k},{\mibs p}}, \\
&&\Lambda^{mn}_{{\mibs k},{\mibs p}}=M^{0m0n}_{{\mibs k},{\mibs p}}.
\end{eqnarray}
If $D^{AB}_{ij}$ is real, then we have the following relations
\begin{equation}
\Omega^{mn}_{{\mibs k},{\mibs p}}=(\Omega^{mn}_{-{\mibs k},-{\mibs p}})^*,\;\;\;
\Lambda^{mn}_{{\mibs k},{\mibs p}}=(\Lambda^{mn}_{-{\mibs k},-{\mibs p}})^*.
\label{olrel}
\end{equation}
In eq.~(\ref{bh0}), the first term denotes the energy of the classical ground state.
The second term linear in boson operators must vanish, since otherwise the assumed classical ground state is no longer stabilized.

To diagonalize the Hamiltonian (\ref{bh0}) with respect to ${\mib k}$, we introduce the reciprocal lattice vectors ${\mib G}_M$ such that the wave vector ${\mib k}$ within the magnetic Brillouin zone plus ${\mib G}_M$ spans the original Brillouin zone.
By noting that $\Omega^{mn}_{{\mibs k},{\mibs p}}$ and $\Lambda^{mn}_{{\mibs k},{\mibs p}}$ are proportional to $\delta_{{\mibs p},{\mibs k}+{\mibs G}_M}$ with $M$ corresponding to the ordering vectors, we obtain the Hamiltonian of eq.~(\ref{boham}).

%
%

%
%
\begin{table}
\caption{The multipole operators in the $\Gamma_8$ subspace.}
\begin{tabular}[t]{@{\hspace{\tabcolsep}\extracolsep{\fill}}cccc} \hline
$\Gamma$ & $\gamma$ & symmetry & $X^A$ \\ \hline
$2u$ &     & $\sqrt{15}xyz$        & $\tau^y$ \\
$3g$ & $a$ & $(3z^2-r^2)/2$        & $\tau^z$ \\
     & $b$ & $\sqrt{3}(x^2-y^2)/2$ & $\tau^x$ \\
$4u1$ & $x$ & $x$                   & $\sigma^x$ \\
      & $y$ & $y$                   & $\sigma^y$ \\
      & $z$ & $z$                   & $\sigma^z$ \\
$4u2$ & $x$ & $x(5x^2-3r^2)/2$      & $\eta^+\sigma^x$ \\
      & $y$ & $y(5y^2-3r^2)/2$      & $\eta^-\sigma^y$ \\
      & $z$ & $z(5z^2-3r^2)/2$      & $\tau^z\sigma^z$ \\
$5u$  & $x$ & $\sqrt{15}x(y^2-z^2)/2$ & $\zeta^+\sigma^x$ \\
      & $y$ & $\sqrt{15}y(z^2-x^2)/2$ & $\zeta^-\sigma^y$ \\
      & $z$ & $\sqrt{15}z(x^2-y^2)/2$ & $\tau^x\sigma^z$ \\
$5g$  & $x$ & $\sqrt{3}yz$            & $\tau^y\sigma^x$ \\
      & $y$ & $\sqrt{3}zx$            & $\tau^y\sigma^y$ \\
      & $z$ & $\sqrt{3}xy$            & $\tau^y\sigma^z$ \\ \hline
\end{tabular}
\end{table}

%
%

\onecolumn

\begin{figure}
\begin{center}
\epsfxsize=15cm \epsfbox{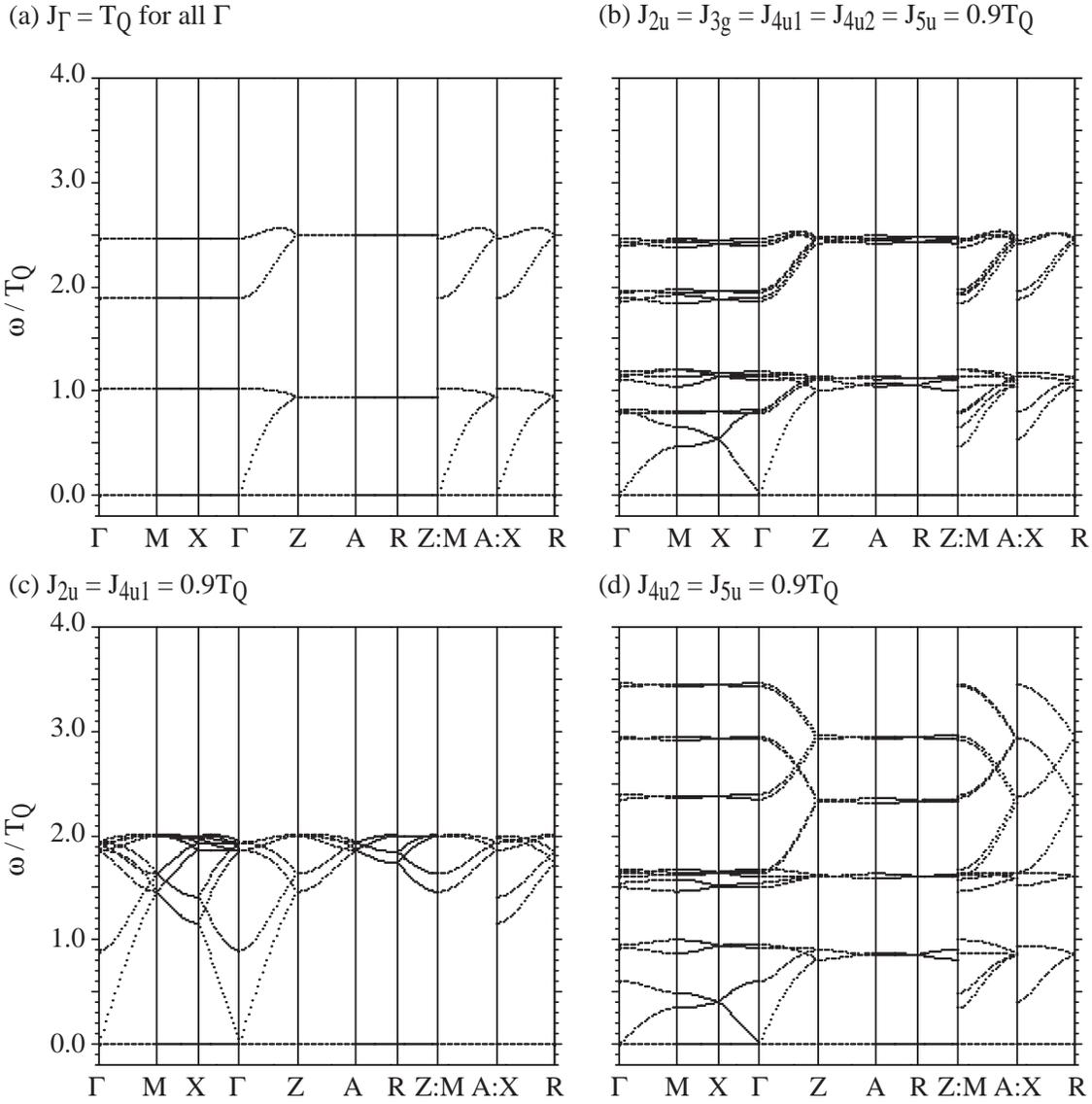}
\end{center}
\caption{The collective excitations without the next-nearest-neighbor interactions. The mean-field solution corresponding to the observed non-collinear ordering is used for the classical ground state.}
\label{dispnn}
\end{figure}

\begin{figure}
\begin{center}
\epsfxsize=15cm \epsfbox{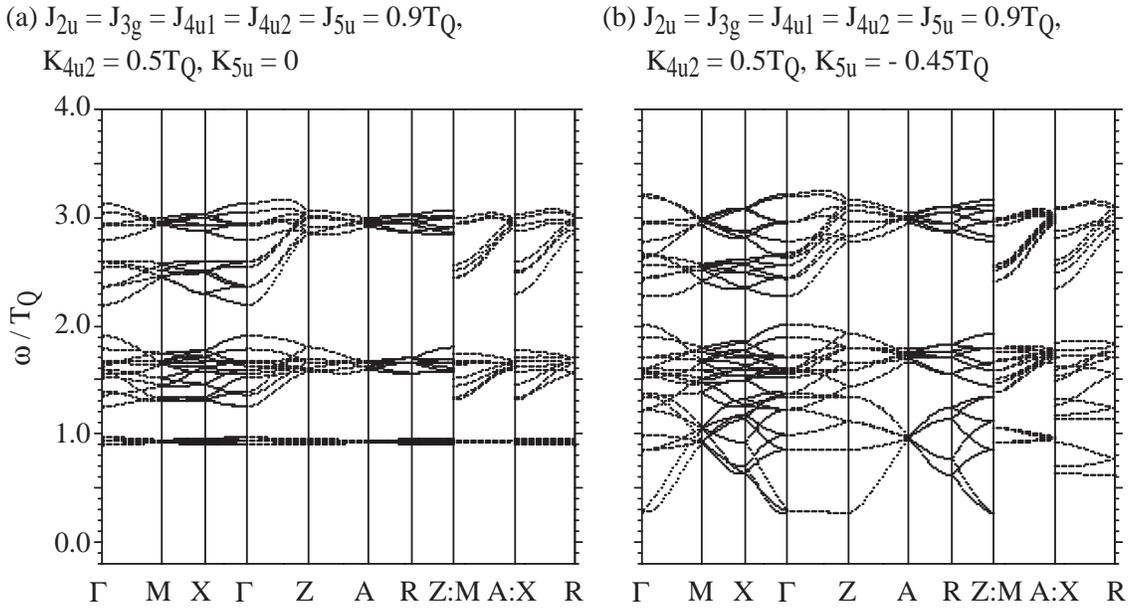}
\end{center}
\caption{The collective excitations for (a) $K_{4u2}=0.5T_Q$ and (b) $K_{4u2}=0.5T_Q$, $K_{5u}=-0.45T_Q$ with the nearest-neighbor interactions close to the SU(4) limit, $J_{2u}=J_{3g}=J_{4u1}=J_{4u2}=J_{5u}=0.9T_Q$.}
\label{dispnnn}
\end{figure}

\end{document}